\documentstyle[amsfonts,preprint,aps]{revtex}
\preprint{\bf Preprint UNO-HEP-98-05}
\begin{document}
\draft
\title{
THE ${\bf N-\Delta}$ WEAK AXIAL-VECTOR
AMPLITUDE ${\bf C_{5}^A(0)}$
}
\author{
Milton Dean Slaughter \footnote{E-Mail address (Internet): mslaught@uno.edu \\To Be Submitted to Modern Physics Letters A}
}

\address{
Department of Physics, University of New Orleans, New Orleans, LA 70148
}
\date{September 1998}

\maketitle

\begin{abstract}
The weak $N-\Delta $ axial-vector transition amplitude $
\left\langle \Delta \right| A_{\pi ^{+}}^\mu \left| N\right\rangle $
---important in $N^{*}$ production processes in general and in isobar models
describing $\nu _\mu N\rightarrow \mu \Delta $ processes in particular
---is examined using a broken symmetry algebraic approach to QCD which
involves the realization of chiral current algebras. We calculate a value for the form
factor $C_{5}^{A}(0)$ in good agreement with experiment.
\end{abstract}


\newpage

\section{Introduction}

The ${\bf N-\Delta }$\ weak axial-vector transition matrix element
is important when one considers: Neutrino quasielastic scattering ($\nu _\mu
+n\rightarrow \mu ^{-}+p$) ; $\Delta ^{++}$\ production reactions ($\nu _\mu
+p\rightarrow \mu ^{-}+\Delta ^{++}$); Hyperon semi-leptonic decays;
Increased understanding of higher symmetries and relativistic and
non-relativistic quark models; Models involving isobars; Dynamical
calculations involving QCD; Dispersion relations; and Current algebras \cite
{chls,dl}.

\section{Theory: The $\Delta ^{++}$\ Production Process}

\noindent The $\Delta ^{++}$\ production process can be studied in our model
by considering the low energy matrix element

\[
\begin{array}{c}
\left\langle \mu ^{-}\Delta ^{++}\left| \nu p\right. \right\rangle =\frac G{%
\sqrt{2}}J_\mu ^hJ_l^\mu =\frac G{\sqrt{2}}\left\langle \Delta ^{++}\right|
V_\mu -A_\mu \left| p\right\rangle J_l^\mu \mbox{, } \\ 
\\ 
\mbox{where }J_\mu ^h\equiv \mbox{hadronic weak current and }J_l^\mu \equiv %
\mbox{leptonic current.} \\ 
V_\mu \;(A_\mu )\mbox{ is the hadronic vector
(axial) current.}
\end{array}
\]

With our normalization,
\footnote{
We normalize physical states according to $<\overrightarrow{p^{\prime }}\,|\,
\overrightarrow{p}>=\delta ^{3}(\overrightarrow{p^{\prime }}\,-%
\overrightarrow{p})$. Dirac spinors are normalized by $\overline{u(p)}u(p)=1$
. Our conventions for Dirac matrices are $\left\{ \gamma ^{\mu },\gamma
^{\nu }\right\} =2g^{\mu \nu }$ with $\gamma _{5}\equiv i\gamma ^{0}\gamma
^{1}\gamma ^{2}\gamma ^{3}$, where $g^{\mu \nu }=$ Diag $[1,-1,-1,-1]$. The
Ricci-Levi-Civita tensor is defined by $\varepsilon _{0123}=-\varepsilon
^{0123}=1=\varepsilon _{123}$.
}
nucleon--nucleon hadronic matrix elements
may be written as:

\begin{equation}
\left\langle B_{2}{(}p_{2},\lambda _{2})\right| J_{\mu }^{h}\left|
B_{1}(p_{1}{,\lambda _{1})}\right\rangle ={\frac{1}{{(2\pi )^{3}}}}\sqrt{{%
\frac{{m}_{1}{m}_{2}}{{E_{1}E_{2}}}}}\bar{u}_{2}(p_{2},\lambda _{2})\left[ {%
\Gamma _{\mu }}\right] u_{1}\left( p_{1},\lambda _{1}\right)  \label{eq1}
\end{equation}

where

\begin{equation}
\begin{array}{ccl}
\Gamma _{\mu } & = & f_{1}(q^{\prime \prime 2})\gamma _{\mu
}+(if_{2}(q^{\prime \prime 2})/m_{1})\sigma _{\mu \nu }q^{\prime \prime
\upsilon }+(if_{3}(q^{\prime \prime 2})/m_{1})q_{\mu }^{\prime \prime } \\ 
& + & \{g_{A}(q^{\prime \prime 2})\gamma _{\mu }+(ig_{P}(q^{\prime \prime
2})/m_{1})\sigma _{\mu \nu }q^{\prime \prime \upsilon }+(ig_{3}(q^{\prime
\prime 2})/m_{1})q_{\mu }^{\prime \prime }\}\gamma _{5}
\end{array}
\label{eq2}
\end{equation}

In Eqs. (\ref{eq1}) and (\ref{eq2} ), $u_1(p_1,\lambda _1)$ is the Dirac
spinor for the inital state (octet) baryon which has mass $m_1$, four
momentum $p_1$, and helicity $\lambda _1$. Similarly, $u_2(p_2,\lambda _2)$,
is the Dirac spinor for the final state (octet) baryon which has mass $m_2$,
four momentum $p_2$, and helicity $\lambda _2$, and $q^{\prime \prime}
\equiv p_2-p_1$.

When the initial baryon is a decuplet state and the final state baryon is a
octet state we have (in the notation of Mathews\cite{mathews}):

\begin{equation}  \label{eq3}
\left\langle B_2{(}p_2,\lambda _2)\right| J_\mu ^h\left| B_1^{\prime }(p_1{%
,\lambda _1)}\right\rangle ={\frac 1{{(2\pi )^3}}}\sqrt{{\frac{{m}%
_1^{^{\prime }}{m}_2}{{E}^{\prime }{_1E_2}}}}\bar u_2(p_2,\lambda _2)\left[ {%
\Gamma _{\mu \beta }}\right] u_1^\beta (p_1,\lambda _1)
\end{equation}

\begin{equation}
\begin{array}{ccl}
\Gamma _{\mu \beta } & = & (f_{1}^{\prime }(q^{2})+g_{1}^{\prime
}(q^{2})\gamma _{5})g_{\mu \beta }+(f_{2}^{\prime }(q^{2})+g_{2}^{\prime
}(q^{2})\gamma _{5})\gamma _{\mu }q_{\beta } \\ 
& + & (f_{3}^{\prime }(q^{2})+g_{3}^{\prime }(q^{2})\gamma _{5})q_{\mu
}q_{\beta }+(f_{4}^{\prime }(q^{2})+g_{4}^{\prime }(q^{2})\gamma
_{5})p_{1\mu }q_{\beta }
\end{array}
\label{eq4}
\end{equation}

Where $u_{1}^{\beta }(p_{1},\lambda _{1})$ is a Rarita-Schwinger spinor and
where the $f_{i}^{\prime }$ and $g_{i}^{\prime }$ are axial-vector and
vector form factors respectively. (Or in the notation of C. H.
Llewellyn-Smith\cite{chls} when we write only the \underline{axial-vector}
part, we obtain):

\begin{equation}
\begin{array}{ccl}
\Gamma _{\mu \beta }^{Axial} & = & C_{5}^{A}(q^{2})g_{\mu \beta } \\ 
& + & C_{6}^{A}(q^{2})q_{\mu }q_{\beta }/m^{2} \\ 
& + & C_{4}^{A}(q^{2})\left\{ (p_{1}\cdot q/m^{2})g_{\mu \beta }+q_{\beta
}(p_{1}+p_{2})_{\mu }/(2m^{2})+q_{\mu }q_{\beta }/(2m^{2})\right\} \\ 
& + & C_{3}^{A}(q^{2})\left\{ ((m^{\ast }-m)/m)g_{\mu \beta }+q_{\beta
}\gamma _{\mu }/m\right\}
\end{array}
\label{eq5}
\end{equation}

The usual Cabibbo assumptions (extended to acknowledge the
existence of $c$, $b$, and $t$ quarks--{\it i.e.} one utilizes the Kobayashi
and Maskawa (K-M) matrix) are then invoked in order to reduce the number of
form factors in Eq.(\ref{eq2} ) from six to four--namely $f_{1}(q^{2})$, $%
f_{2}(q^{2})$, $g_{1}(q^{2}),$and $g_{P}(q^{2})$.

These assumptions are:

\begin{itemize}
\item  Universality of the coupling of the leptonic current to hadronic
current;

\item  $J_\mu ^h$ components transform like the charged members of the $%
SU(3) $ $J^P=0^{-}$ octet;

\item  Generalized CVC (isotriplet current hypothesis) holds---implies that $%
f_{1}$ and $f_{2}$ can be calculated from the known proton and neutron
electromagnetic form factors and that $f_{3}=0$;

\item  No second class currents exist---implies that $g_3=0.$
\end{itemize}

With those assumptions, Eq.(\ref{eq2} ) then effectively reduces to:

\begin{equation}  \label{eq6}
\begin{array}{ccl}
\Gamma _\mu & = & f_1(q^{\prime \prime 2})\gamma _\mu +(if_2(q^{\prime
\prime 2})/m_1)\sigma _{\mu \nu }^\nu q^{\prime \prime }+ \\ 
& + & g_A(q^{\prime \prime 2})\gamma _\mu \gamma _5+(ig_P(q^{\prime \prime
2})/m_1)\sigma _{\mu \nu }q^{\prime \prime \upsilon }\gamma _5.
\end{array}
\end{equation}

For example the well known nucleon weak axial-vector form factor $
g_A(q^{\prime \prime ^{\,2}})$ can be parametrized by

\[
g_A(q^{\prime \prime ^{\,2}})/g_A(0)\cong \left[ 1-q^{\prime \prime
^{\,2}}/m_A^2\right] ^{-2}, 
\]

where

\[
\left\langle p,p_2\right| A_{\pi ^{+}}^\mu (0)\left| n,p_1\right\rangle
\approx (2\pi )^{-3}\sqrt{(mm_n)/(E_{p_1}E_{p_2})}\bar u_p(p_2)\left[
g_A(q^{\prime \prime ^{\,2}})\gamma ^\mu \gamma _5\right] u_n(p_1), 
\]

$m_n=$ neutron mass, and $q^{\prime \prime ^{\,2}}=(p_2-p_1)^2$.

\section{Previous Results and Methodology}

\noindent \sloppy{\ We consider helicity states with $\lambda =+{1/2}$ (i.e.
spin non-flip sum rules) and the non-strange ($S=0$) $L=0$ ground state
baryons ($J^{PC}=\frac{1}{2}^{+},\frac{3}{2}^{+}$). It is well-known that if
one defines the axial-vector matrix elements:}

$\,$%
\[
\left\langle p,1/2\right| A_{\pi ^{+}}\left| n,1/2\right\rangle \equiv
f=g_A(0), 
\]

$\ $%
\[
\left\langle {\Delta ^{++},1/2}\right| A_{\pi ^{+}}\left| {\Delta ^{+},1/2}%
\right\rangle \equiv -\sqrt{{\frac 32}}\,g, 
\]

\[
\left\langle {\Delta ^{++},1/2}\right| A_{\pi ^{+}}\left| {p,1/2}%
\right\rangle \equiv -\sqrt{6}\,h,\ 
\]

and applies asymptotic level realization to the chiral $SU(2)\otimes SU(2)$
charge algebra $\left[ {A_{\pi ^{+}},A_{\pi ^{-}}}\right] =2V_3$, then

\[
h^2={(4/25)}f^{\;2}\ (the\ sign\ of\ h=+(2/5)f)\ \quad and\qquad \ g=(-\sqrt{%
2}/5)f. 
\]

If one further defines (suppressing the index $\mu $ ):

\[
<\,\Delta ^{+},1/2,\vec s\,|\,j_3|\,\,\Delta ^{+},1/2,\vec t>\equiv a,\
\qquad \qquad <p,1/2,\vec s\,|j_3|\,p,1/2,\vec t>\equiv b, 
\]

$\ $%
\[
<n,1/2,\vec s\,|\,j_3|\,\Delta ^0,1/2,\vec t>\equiv c,\qquad \qquad \
<\Delta ^0,1/2,\vec s\,|\,j_3|\,n,1/2,\vec t>\equiv d\ 
\]

\noindent (note that other required matrix elements of $j_3$ can then be
obtained easily from the double commutator $[\,[j_3^\mu (0),\,V_{\pi
^{+}}]\,,V_{\pi ^{-}}]=2\;j_3^\mu (0))$

\noindent and if one also inserts the algebra $\left[ j_3^\mu (0){,}A_{\pi
^{+}}\right] =A_{\pi ^{+}}^\mu {(0)}$ ($j^\mu \equiv j_3^\mu +j_S^\mu $ ,
where $j_3^\mu \equiv $ isovector part of $j^\mu $ and $j_S^\mu $ is
isoscalar) between the ground states $\left\langle B(\alpha ,\lambda =1/2,%
\overrightarrow{s})\right| $and $\left| B^{\prime }(\alpha ,\lambda =1/2,%
\overrightarrow{t})\right\rangle $ with $\left| \overrightarrow{s}\right|
\rightarrow \infty ,\left| \overrightarrow{t}\right| \rightarrow \infty $,
where $\left\langle B(\alpha )\right| $ and $\left| B^{\prime }(\beta
)\right\rangle $ are the following $SU_F(2)$ related combinations:\quad {\ $%
\left\langle p,n\right\rangle ,\ \left\langle p,\Delta ^0\right\rangle $,
\qquad $\left\langle \Delta ^{++},p\right\rangle ,\ \left\langle n,\Delta
^{-}\right\rangle $, \qquad $\left\langle \Delta ^{++},\Delta
^{+}\right\rangle ,\ \left\langle {\Delta ^{+}},{\Delta ^0}\right\rangle $,
\qquad $\left\langle \Delta ^0,\Delta ^{-}\right\rangle \ ,\ \left\langle
\Delta ^{+},n\right\rangle $ } \ , \quad then one obtains (we use $%
<N|\,j_S^\mu |\,\Delta >=0$ ) the constraint equations (not all independent):

\begin{equation}  \label{eq7}
2fb-\sqrt{2}h(c+d)=f^{L=0}(\lambda =1/2)\left\langle {p}\right| A_{\pi
^{+}}^\mu \left| {n}\right\rangle ,
\end{equation}

\begin{equation}  \label{eq8}
\sqrt{2}h(a+b)+(-\sqrt{2}g-f)c=f^{L=0}(\lambda =1/2)\left\langle {p}\right|
A_{\pi ^{+}}^\mu \left| {\Delta }^0\right\rangle ,
\end{equation}

\begin{equation}  \label{eq9}
\sqrt{6}h(-3a+b)+\sqrt{3/2}gd=f^{L=0}(\lambda =1/2)\left\langle {\Delta }%
^{++}\right| A_{\pi ^{+}}^\mu \left| p\right\rangle ,
\end{equation}

\begin{equation}  \label{eq10}
\sqrt{6}h(3a-b)-\sqrt{3/2}gc=f^{L=0}(\lambda =1/2)\left\langle {n}\right|
A_{\pi ^{+}}^\mu \left| {\Delta }^{-}\right\rangle ,
\end{equation}

\begin{equation}  \label{eq11}
-\sqrt{6}ga+\sqrt{6}hc=f^{L=0}(\lambda =1/2)\left\langle {\Delta }%
^{++}\right| A_{\pi ^{+}}^\mu \left| {\Delta }^{+}\right\rangle ,
\end{equation}

\begin{equation}  \label{eq12}
-2\sqrt{2}ga+\sqrt{2}h(c+d)=f^{L=0}(\lambda =1/2)\left\langle {\Delta }%
^{+}\right| A_{\pi ^{+}}^\mu \left| {\Delta }^0\right\rangle ,
\end{equation}

\begin{equation}  \label{eq13}
-\sqrt{6}ga+\sqrt{6}hd=f^{L=0}(\lambda =1/2)\left\langle {\Delta }^0\right|
A_{\pi ^{+}}^\mu \left| {\Delta }^{-}\right\rangle ,
\end{equation}

\begin{equation}  \label{eq14}
-\sqrt{2}h(a+b)+(f+\sqrt{2}g)d=f^{L=0}(\lambda =1/2)\left\langle {\Delta }%
^{+}\right| A_{\pi ^{+}}^\mu \left| {n}\right\rangle .
\end{equation}

Applying asymptotic level symmetry, Eqs. (\ref{eq7})--(\ref{eq14})
immediately imply that 
\begin{equation}  \label{eq15}
d=c
\end{equation}
and 
\begin{equation}  \label{eq16}
a=b+[-\frac 14\frac gh-\frac 1{2\sqrt{2}}\frac fh]c.
\end{equation}
One can calculate $f^{L=0}(\lambda =1/2)$ easily by setting $\mu =0$ ,
restoring the $x$ dependence to the matrix elements and integrating over $%
d\vec x$.

We find that $f^{L=0}(\lambda =1/2)=1$. Eqs. (\ref{eq7})--(\ref{eq14}) then
relate the weak matrix elements $\left\langle {\Delta }^{+},1/2,\vec{s}%
\right| A_{\pi ^{+}}^{\mu }(0)\left| {\Delta }^{0},1/2,\vec{t}\right\rangle $%
, $\left\langle {\Delta }^{+},1/2,\vec{s}\right| A_{\pi ^{+}}^{\mu
}(0)\left| {n},1/2,\vec{t}\right\rangle $, and $\left\langle {p},1/2,\vec{s}%
\right| A_{\pi ^{+}}^{\mu }(0)\left| {\Delta }^{0},1/2,\vec{t}\right\rangle $
to the matrix elements $\left\langle p,1/2,\vec{s}\,|j_{3}^{\mu }(0)|\,p,1/2,%
\vec{t}\right\rangle $ and $\left\langle {p},1/2,\vec{s}\right| A_{\pi
^{+}}^{\mu }(0)\left| {n},1/2,\vec{t}\right\rangle $.

In fact, one discovers the important relation 
\begin{equation}
(8h-2f)b=2\sqrt{2}\left\langle {p}\right| A_{\pi ^{+}}^{\mu }\left| {\Delta }%
^{0}\right\rangle -\left\langle {p}\right| A_{\pi ^{+}}^{\mu }\left| {n}%
\right\rangle .  \label{eq17}
\end{equation}

\section{Results}

\noindent We now choose $\mu =0$ and take the limit $\left| \overrightarrow{s}\right| \rightarrow \infty $
and $\left| \overrightarrow{t}\right| \rightarrow \infty $ ($\overrightarrow{%
s}$ and $\overrightarrow{t}$ are both taken along the $z$-axis), then $%
q^2=q^{^{^{\prime \prime }}2}=\widetilde{q}^2=0$.  We find that

\begin{equation}  \label{eq18}
\begin{array}{rl}
(2\pi )^3<p,1/2,\vec s\,|j_3^0|\,p,1/2,\vec t>\rightarrow & \left[ 1- 
\widetilde{q}^2/(4m^2)\right] ^{-1}G_E^V(\widetilde{q}^2) \\ 
(2\pi )^3\left\langle {p}\right| A_{\pi ^{+}}^0\left| {n}\right\rangle
\rightarrow & g_A(q^{^{^{\prime \prime }}2}) \\ 
(2\pi )^3\left\langle {p}\right| A_{\pi ^{+}}^\mu \left| {\Delta }%
^0\right\rangle \rightarrow & \sqrt{\frac 23}\frac{(m+m^{*})}{2m^{*}}%
C_5^A(q^2)
\end{array}
\end{equation}

Thus, Eq. (\ref
{eq17}) and Eq. (\ref{eq18}) then predict that

\begin{equation}  \label{eq19}
C_5^A(0)=\frac 45\sqrt{3}\frac{m^{*}}{(m+m^{*})}g_A(0).
\end{equation}

Numerically Eq. (\ref{eq19}) reads (using $g_A(0)=1.25$, $m^{*}=1.232 $ $%
GeV/c^2$)

\begin{equation}  \label{eq20}
C_5^A(0)=0.98
\end{equation}

This value is consistent with PCAC and yields the value $\Gamma (\Delta
)\approx 100$ MeV.

\section{Conclusions and Some Comparisons with Other Models and Experiment}

\begin{itemize}
\item  $SU(6)$ Symmetry Predicts
\end{itemize}

\[
C_5^A(q^2=0)^{n\rightarrow \Delta ^{+}}=\frac 25\sqrt{3}\left(
g_A/g_V\right) ^{n\rightarrow p}. 
\]

This gives rise to the following dilemma:
Does one use the pure $SU(6)$ result $\left( g_A/g_V\right) ^{n\rightarrow
p}=5/3$ $\Longrightarrow ${\ $C_A^5(q^2=0)^{n\rightarrow \Delta ^{+}}=1.15$ or does one use the experimental value of $\left( g_A/g_V\right)
^{n\rightarrow p}=1.25\Longrightarrow $ $C_5^A(q^2=0)^{n\rightarrow \Delta
^{+}}=0.87$ ? }

Clearly, the value of $C_5^A(q^2=0)^{n\rightarrow \Delta ^{+}}$ that one chooses to
use can represent almost a factor of two in the predicted value of $\Gamma
(\Delta )$.

\begin{itemize}
\item  Non-Relativistic Conventional ($N$ and $\Delta $ space wave functions
are identical) Quark Model
\end{itemize}

\[
C_5^A(q^2=0)^{n\rightarrow \Delta ^{+}}=\frac 25\sqrt{3}\left( g_A\right)
^{n\rightarrow p}=0.87. 
\]

\begin{itemize}
\item  Static (Yukawa pion-nucleon coupling ) Models
\end{itemize}

\[
C_5^A(q^2=0)^{n\rightarrow \Delta ^{+}}=\frac{g_{\Delta ^{++}p}}{\sqrt{6}g_N}%
\left( g_A\right) ^{n\rightarrow p}=1.11. 
\]

\begin{itemize}
\item  Adler's Model and the Feynman, Kislinger, Ravndal (FKR) Relativistic
Quark Model
\end{itemize}

The predictions of the Adler and FKR models are quite similar. Both models predict (roughly)
the same non-relativistic limits with the $q^2$ dependence of axial-vector transition
amplitudes determined by $q^2$ dependence of nucleon axial-vector form
factors, although the FKR model makes the stronger prediction that the dependence is
the same. It is also true that in Adler and FKR models the low $q^2$ dependence is roughly the same as for the Static model.

\begin{itemize}
\item  Experiment (CERN, Brookhaven, Argonne: by measuring the axial-vector
mass $M_A$ ) generally favors the Adler model.
\end{itemize}

\begin{itemize}
\item  PCAC
\end{itemize}

From the process $\upsilon n\rightarrow \Delta ^{+}$ ({\it i.e.} $\upsilon
+n\rightarrow \mu ^{-}+\Delta ^{+}$), PCAC predicts that

\[
C_5^A(q^2=0)^{n\rightarrow \Delta ^{+}}=\frac{f_\pi g_{\Delta ^{+}n\pi ^{+}}}%
m=1.2, 
\]

If the Goldberger-Treiman relationship $g_A/g_N=f_\pi /\sqrt{2}m $ is
exactly satisfied, then the static model prediction and PCAC predictions are
the same.

\begin{itemize}
\item  We conclude that our broken symmetry algebraic approach to the
calculation of $C_5^A(q^2=0)^{n\rightarrow \Delta ^{+}}$ yields results
consistent not only with experiment but also with the widely used Adler model. The broken
symmetry approach also correctly gives the $\Delta $ width, and resolves mass and wave function degeneracy  problems present in many widely-used quark models.
\end{itemize}

\end{document}